\documentclass[12pt,twoside]{article}
\usepackage{fleqn,epsf,espcrc1}
\usepackage{latexsym,amsfonts}

\usepackage{graphicx}
\usepackage[figuresright]{rotating}

\newcommand{\AmS}{{\protect\the\textfont2
  A\kern-.1667em\lower.5ex\hbox{M}\kern-.125emS}}

\title{Chemical freeze-out parameters at RHIC from microscopic
       model calculations }

\author{
L.~V.~Bravina$^a$, E.~E.~Zabrodin$^{a,b}$,
S.~A.~Bass$^{c,d}$,
A.~Faessler$^b$, C.~Fuchs$^b$,
M.~I.~Gorenstein$^{e,f}$, 
W.~Greiner$^e$, S.~Soff$^g$, H.~St{\"o}cker$^e$, H.~Weber$^e$
\\
\vspace*{.25 cm}
{\small\it
$^a$Institute for Nuclear Physics, Moscow State University,
119899 Moscow, Russia} \\
{\small\it
$^b$Institute for Theoretical Physics, University of
T\"ubingen, D-72076 T\"ubingen, Germany} \\
{\small\it
$^c$Department of Physics, Duke University, Durham NC 27708, USA}\\
{\small\it
$^d$RIKEN-BNL Research Center, Brookhaven National Laboratory,
Upton, NY 11973, USA} \\
{\small\it
$^e$Institute for Theoretical Physics, University of Frankfurt,
D-60054 Frankfurt, Germany} \\
{\small\it
$^f$Bogolyubov Institute for Theoretical Physics, Kiev, Ukraine} \\
{\small\it
$^g$Nuclear Science Division, Lawrence Berkeley National Laboratory, 
CA 94720, USA}
}
       
\begin{document}

\maketitle

\vspace*{-.3cm}

\begin{abstract}
{\small
The relaxation of hot nuclear matter to an equilibrated state in the
central zone of heavy-ion collisions at energies from AGS to RHIC is 
studied within the microscopic UrQMD model. It is found that the 
system reaches the (quasi)equilibrium stage for the period of 10-15 
fm/$c$. Within this time the matter in the cell expands nearly
isentropically with the entropy to baryon ratio $S/A = 150 - 170$.
Thermodynamic characteristics of the system at AGS and at SPS 
energies at the endpoints of this stage are very close to the 
parameters of chemical and thermal freeze-out extracted from the 
thermal fit to experimental data. Predictions are made for the full 
RHIC energy $\sqrt{s} = 200$ AGeV. The formation of a resonance-rich 
state at RHIC energies is discussed.
}
\end{abstract}

\vspace{.3 cm}

\begin{figure}[htb]
\begin{minipage}[t]{69mm}
\vspace*{-0.8cm}
\centerline{\epsfysize=75mm\epsfbox{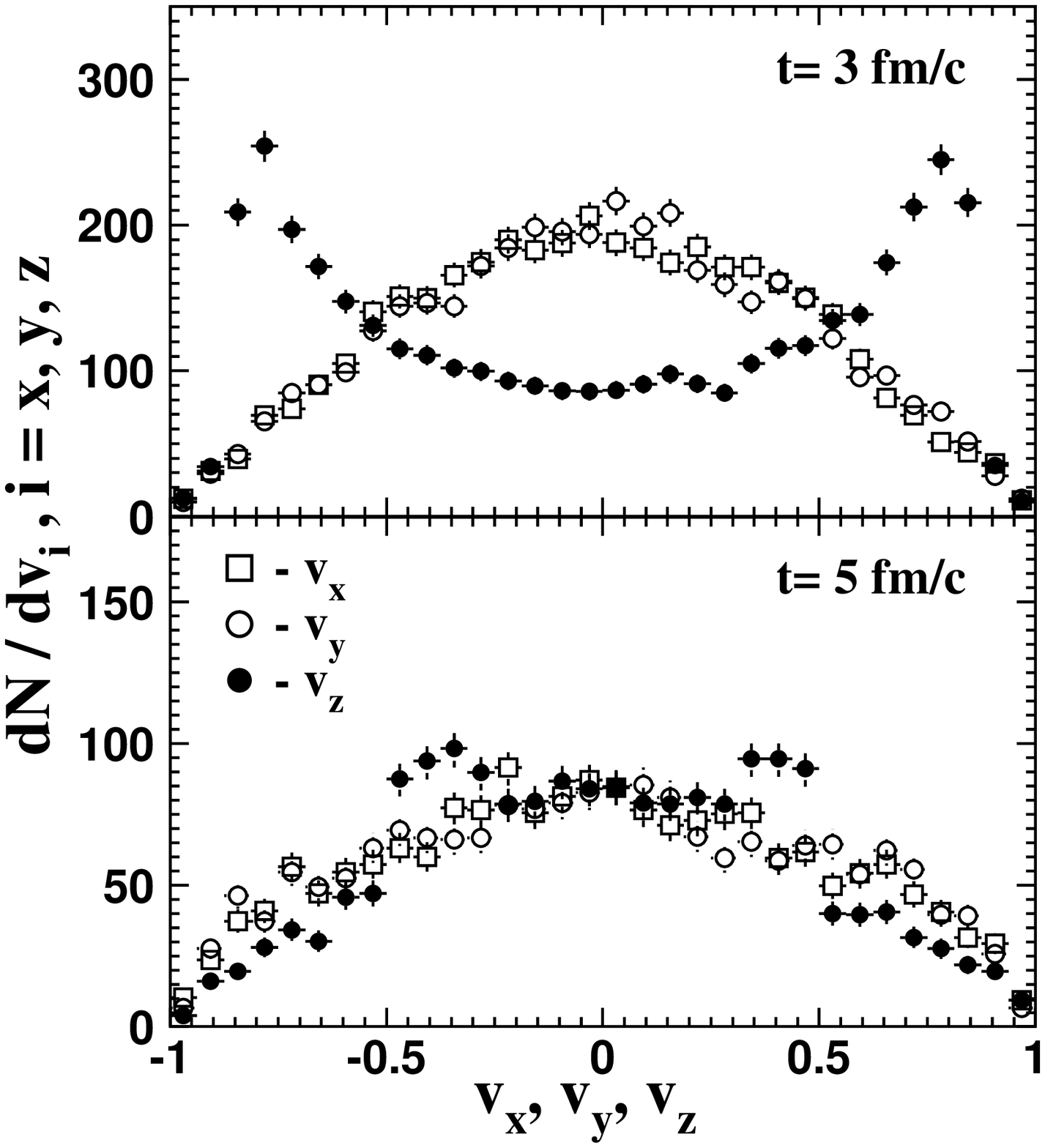}}
\vspace*{-0.9cm}
\caption{\small 
Hadron velocity distributions $d N / d v_i$, $i = x$ ($\Box$), 
$y$ ($\bigcirc$), and $z$ ($\bullet$) at $t = 3$ fm/$c$ and 
$t= 5$ fm/$c$ in a central cell in Au+Au collisions at RHIC.
 }
\label{fig1}
\end{minipage}
\hspace{\fill}
\begin{minipage}[t]{83mm}

\vspace*{-0.7cm}
\centerline{\epsfysize=75mm\epsfbox{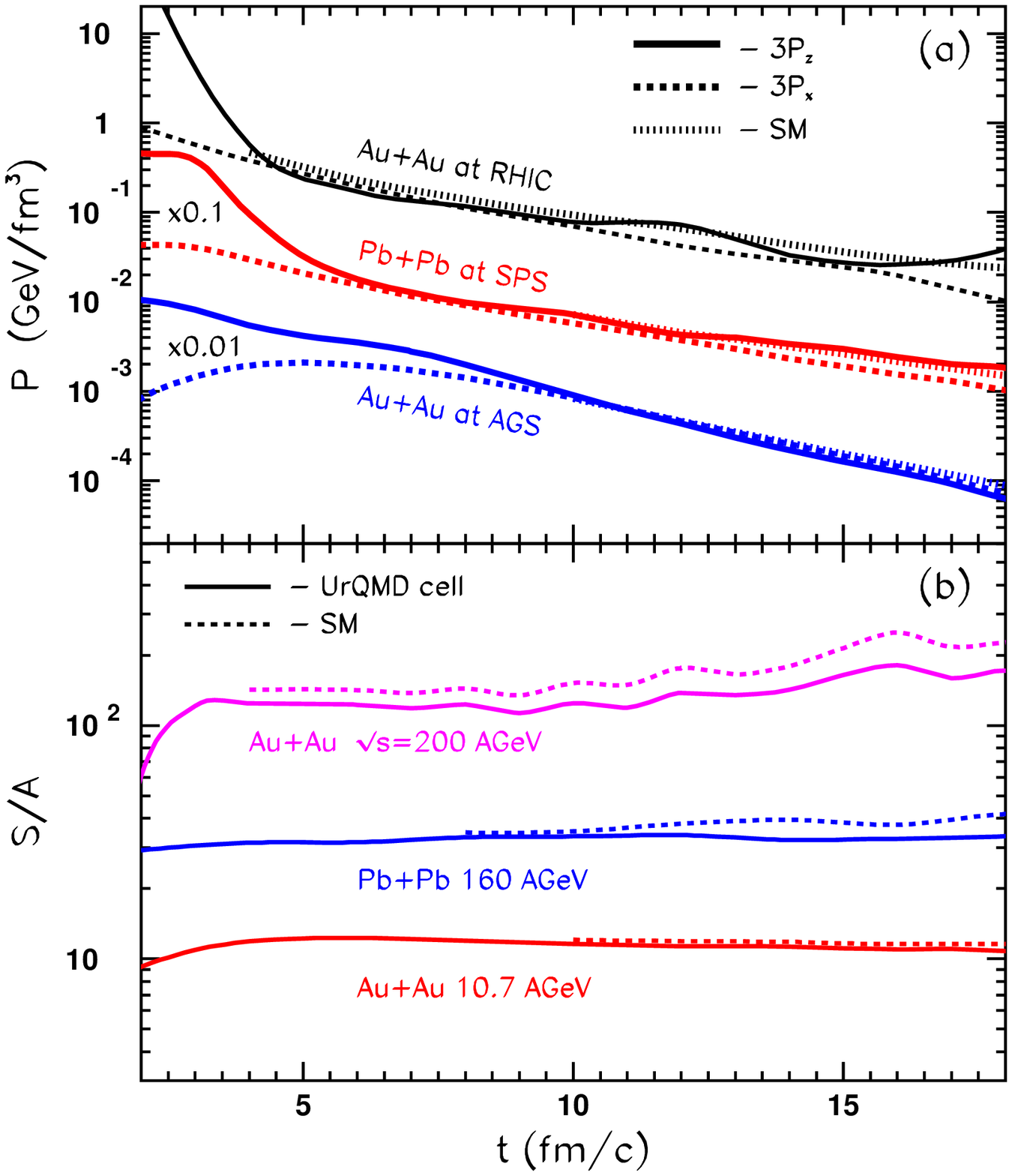}}
\vspace*{-0.9cm}
\caption{\small 
Time evolution of the components of the pressure tensor $P$ (a) and
of the entropy per baryon $S/A$ (b) in the central cell of heavy 
ion collisions compared to the SM results.
 }
\label{fig2}
\end{minipage}
\end{figure}

Thermalization and chemical equilibration of hot and dense nuclear
matter produced in ultrarelativistic heavy-ion collisions is
a topic of great importance for the interpretation of current
SPS and RHIC results.
In our investigation we study the relaxation process in
central Au+Au collisions at $\sqrt{s} = 200$ AGeV within the
microscopic transport UrQMD model \cite{urqmd}.
Earlier studies at AGS and SPS energies revealed that the central
reaction volume defined by a
cubic cell of volume $V = 125$ fm$^3$ is well suited for this kind of
study \cite{loceq}. 
It contains enough particles to be treated as a statistical system,
and its macroscopic characteristics become isotropic after some
time. 
Figure~\ref{fig1} depicts the
velocity distributions of hadrons in the cell in transverse ($x$
and $y$) and in longitudinal ($z$) directions. At $t = 3$ fm/$c$
the longitudinal velocity distribution differs considerably from
the distributions in the transverse plane, while at $t = 5$ fm/$c$ the 
magnitudes and widths of all three distributions become very close
to each other. Isotropy of the velocity distributions results in
the isotropy of pressure in the cell. Pressure in longitudinal
and in transverse direction is shown in Fig.~\ref{fig2}(a) for
AGS, SPS, and RHIC energies. It is widely believed that the 
thermalization at RHIC sets in quite early.
Indeed, at RHIC the pressure in the cell becomes isotropic at
$t \approx 5$ fm/$c$ compared with $t \approx 8$ fm/$c$ (SPS) and
$t \approx 10$ fm/$c$ (AGS).
Starting from $t = 5$ fm/$c$ the results of the microscopic 
calculations are compared with the predictions of the statistical
model (SM) of an ideal hadron gas \cite{SM}. The values of the
energy density $\varepsilon$, baryon density $\rho_B$, and
strangeness density $\rho_S$, determined microscopically, are used
as an input to obtain particle yields, partial energy densities,
pressure, and entropy density via the temperature $T$, baryochemical
potential $\mu_B$, and strangeness chemical potential $\mu_S$.
As seen in Fig.~\ref{fig2}(a), the microscopic pressure is very
close to the grand canonical pressure after the onset of the 
(quasi)equilibrium stage.

\vspace*{-0.85cm}
The entropy density per baryon $S/A$, defined both microscopically
and macroscopically (via the Gibbs thermodynamic identity $Ts = 
\varepsilon + P - \mu_B \rho_B -\mu_S \rho_S$) is presented in
Fig.~\ref{fig2}(b). The expansion in the cell proceeds nearly
isentropically with $S/A = 12$ (AGS), 32 (SPS), and 150 (RHIC)
(cf. $(S/A)^{\rm therm.fit} \cong 14$ (AGS) and $\cong 36$ (SPS) 
\cite{ClRe99}). Therefore, it would be very interesting to compare
the UrQMD estimate $(S/A)^{\rm RHIC} = 150 - 170$ with the value 
extracted from the thermal model fit to RHIC experimental data. 

\begin{figure}[htb]
\begin{minipage}[t]{66mm}
\vspace*{-1.5cm}
\centerline{\epsfysize=75mm\epsfbox{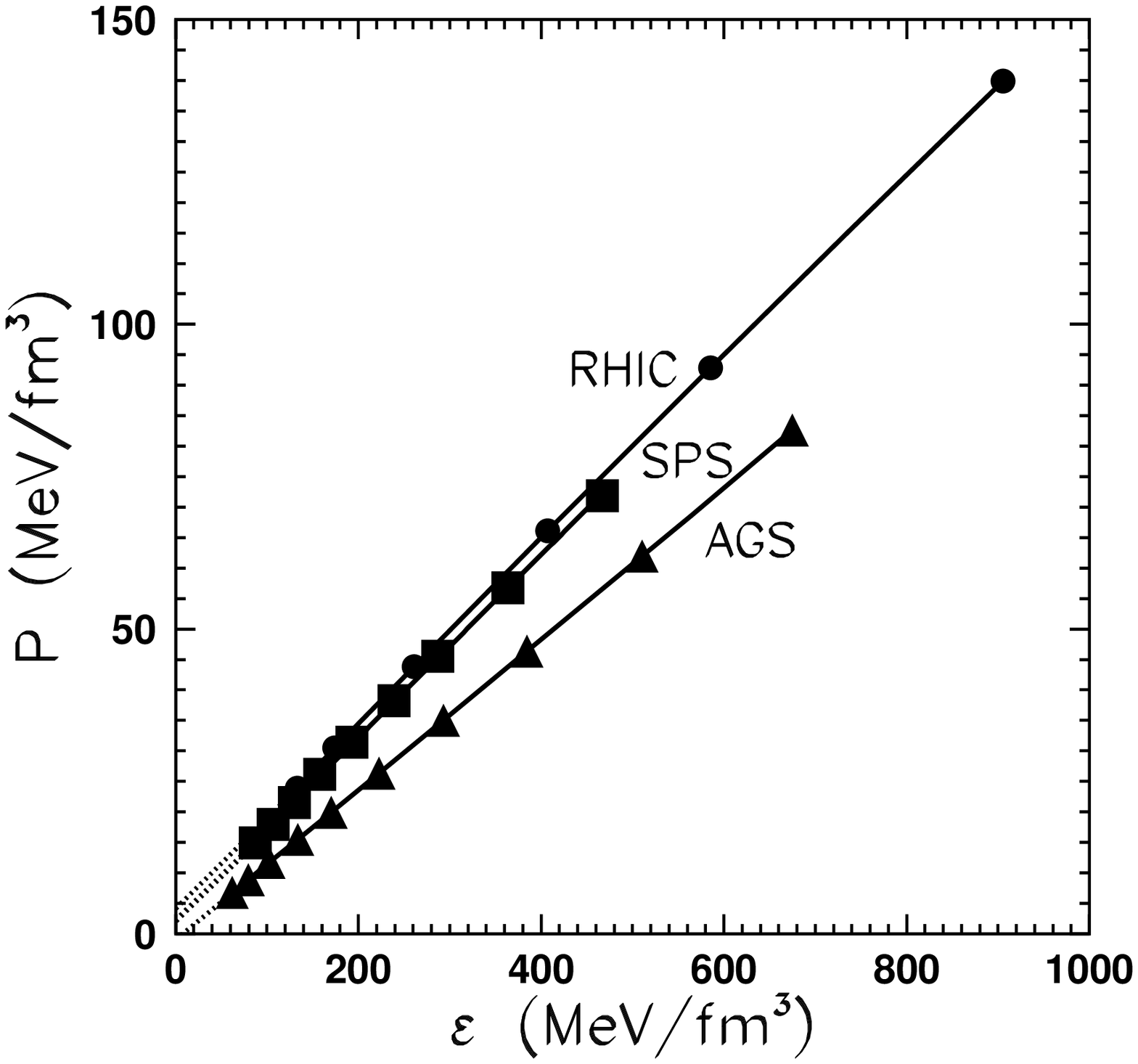}}
\vspace*{-1.2cm}
\caption{\small 
The evolution of pressure $P$ and baryon density
$\varepsilon$ in the central cells of the heavy-ion collisions
at AGS, SPS, and RHIC energies.
 }
\label{fig3}
\end{minipage}
\hspace{\fill}
\begin{minipage}[t]{87mm}

\vspace*{-1.5cm}
\centerline{\epsfysize=75mm
\epsfbox{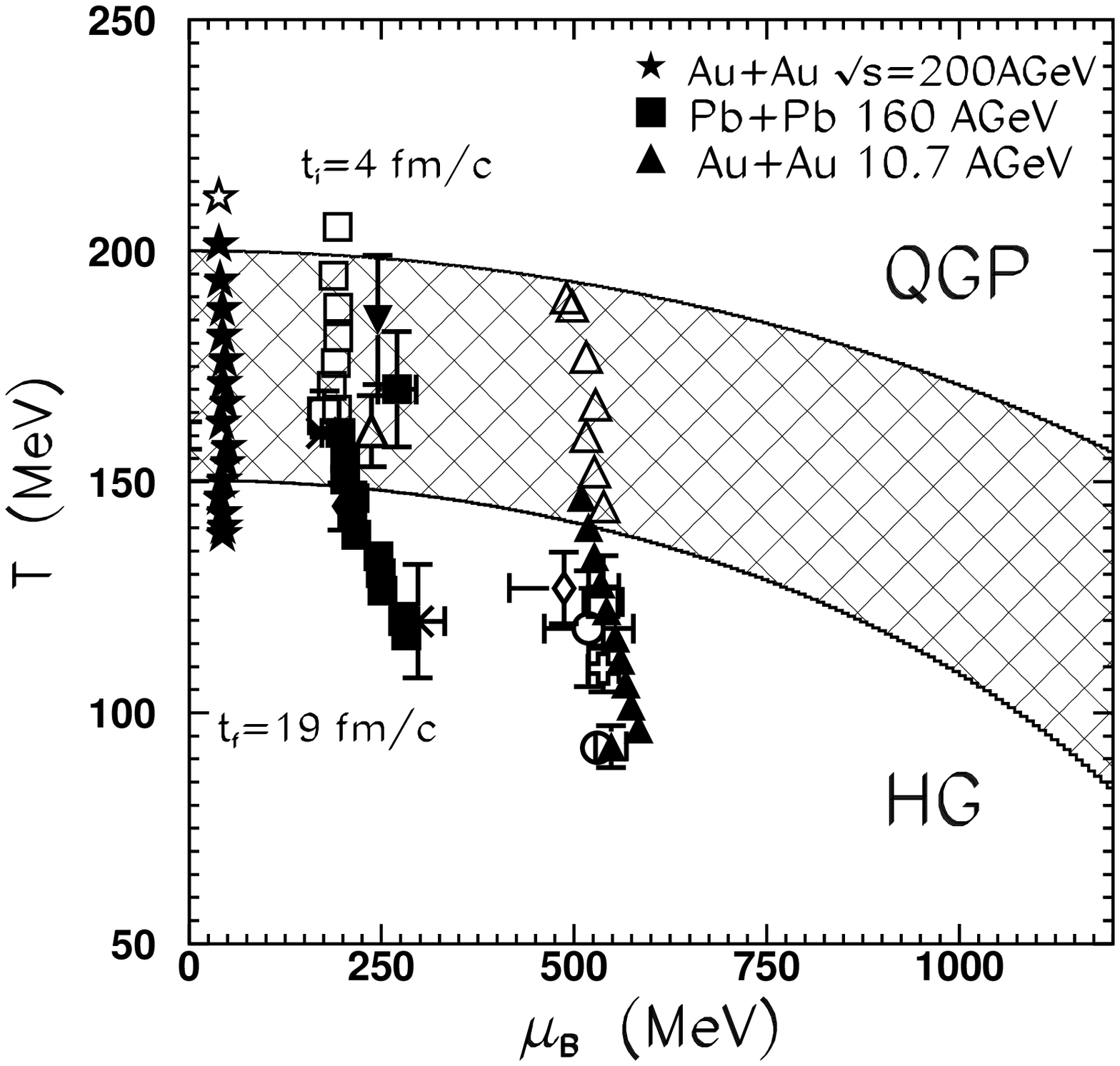}}
\vspace*{-1.4cm}
\caption{\small 
The same as Fig.3 but for the $(T, \mu_{\rm B})$-plane.
Solid symbols correspond to the stage of kinetic equilibrium,
open symbols indicate the preequilibrium stage. Symbols with error 
bars are chemical and thermal freeze-out parameters obtained from 
the SM fit \cite{ClRe99}. The hatched area shows the expected region 
of the quark-hadron phase transition.
 }
\label{fig4}
\end{minipage}
\end{figure}

The equation of state (EOS) in the $(P, \varepsilon)$-plane is
shown in Fig.~\ref{fig3}. For all three energies it can be 
well approximated by a simple linear dependence 
$P/\varepsilon = 0.12$ (AGS), and 0.15 (SPS and RHIC). Note, that
this version of the model does not imply the formation of the
quark-gluon plasma, therefore, there are no kinks in this plot that
can be attributed to quark-hadron phase transition. The evolution of 
the EOS in the 
$(T,\mu_B)$-plane is depicted in Fig.~\ref{fig4} together with the 
chemical freeze-out and the thermal freeze-out parameters obtained
from a thermal fit to experimental data at AGS and SPS energies.  
One can see that temperatures and chemical potentials in the cells
at the beginning and at the end of the equilibrated stage are close
to the thermodynamic parameters of chemical and thermal freeze-out,
respectively. The UrQMD predicts that at chemical freeze-out
$\displaystyle T_{\rm chem.FO}^{\rm RHIC} = 195 \pm 5$ MeV and 
$\displaystyle \mu_{\rm chem.FO}^{\rm RHIC} = 43 \pm 2$ MeV, while at 
thermal freeze-out $\displaystyle T_{\rm therm. FO}^{\rm RHIC} = 130 
\pm 5$ MeV and $\displaystyle \mu_{\rm therm. FO}^{\rm RHIC} = 43 \pm 
3$ MeV, i.e., the evolution of nuclear matter in the cell proceeds at 
constant baryon chemical potential. Calculations show very weak 
(within few MeV) difference between the results for Au+Au at $\sqrt{s} 
= 130$ AGeV and at $\sqrt{s} = 200$ AGeV. A thermal fit to particle 
ratios measured by the STAR Collaboration (preliminary data) yields 
$T_{\rm chem.FO} = 190 \pm 20$ MeV and $\mu_{q} = 15 \pm 5$ MeV 
\cite{nuxu}, i.e. $\mu_B = 45 \pm 15$ MeV, which is in remarkable 
agreement with the UrQMD calculations.

\vspace*{-0.85cm}

It is interesting to check the correspondence of the
cell conditions at chemical freeze-out to the criterion
$\langle E \rangle /\langle N_h \rangle \approx \langle M_h \rangle +
3/2 T \approx 1$ GeV introduced in \cite{ClRe99}.\, 
The\, mean\, energy

\begin{table}[htb]
\begin{minipage}[t]{75mm}
\vspace*{-0.8cm}
\caption{\small Energy per hadron $\langle E \rangle /\langle N_h 
\rangle $, energy density $\varepsilon$, and temperature $T$ in the 
cell at the beginning of kinetic equilibrium.
 }
\label{tab1}
\begin{tabular}{@{}ccccc}
\hline\hline
 & time  & $\varepsilon$ & $  \langle E \rangle / \langle N_h 
\rangle $ & $ T $ \\ &fm/$c$ &  GeV/fm$^3$   &  GeV & MeV    \\
\hline\hline
AGS    & 10  & 0.68  & 1.11 & 129   \\
AGS    & 12  & 0.39  & 1.06 & 116   \\
SPS    & 8   & 0.74  & 0.88 & 170   \\
SPS    & 10  & 0.46  & 0.80 & 161   \\
RHIC   & 5   & 2.33  & 1.08 & 201   \\
RHIC   & 6   & 1.70  & 1.01 & 193   \\
\hline\hline
\end{tabular}
\end{minipage}
\hspace{\fill}
\begin{minipage}[t]{73mm}
\vspace*{-0.85cm}
per particle at the beginning of the equilibration in the cell at
AGS, SPS, and RHIC energies is listed in Table~\ref{tab1}.
With rising bombarding energy from AGS to SPS there is a transition 
from baryon to meson dominated matter.
It leads to the drop of $\langle M_h \rangle$ from nucleon mass to 
mass of $\rho$ meson in accord with \cite{ClRe99}. With further
increase of the freeze-out temperature the yields of heavy meson 
resonances rise faster than that of light mesons, thus leading to
the rise of $\langle M_h \rangle$ with $\sqrt{s}$.
\end{minipage}
\end{table}
\vspace*{-0.5cm}

\begin{figure}[htb]
\begin{minipage}[t]{69mm}
\vspace*{-0.9cm}
\centerline{\epsfysize=79mm\epsfbox{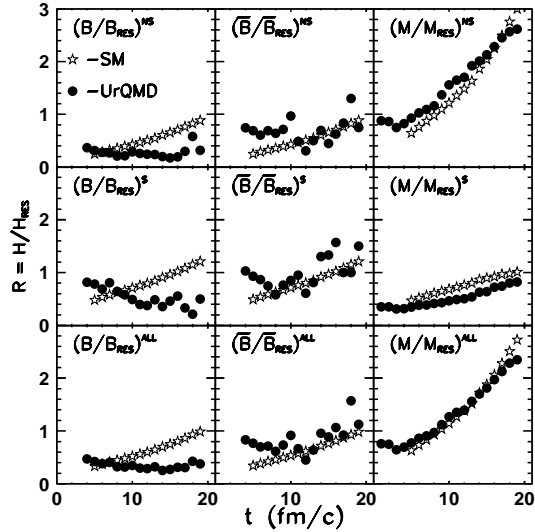}}
\vspace*{-0.6cm}
\caption{\small 
Time evolution of the hadron to resonance ratio $R = H/H_{res}$ in
the central cell of Au+Au collisions at RHIC shown separately for
baryons, antibaryons, and mesons as well as for non-strange hadrons,
strange hadrons, and total hadron yields. Circles denote the UrQMD 
predictions, stars correspond to the SM results.
 }
\label{fig5}
\end{minipage}
\hspace{\fill}
\begin{minipage}[t]{80mm}
\vspace*{-0.88cm}
hot hadronic matter in 
the cell as well as in the whole volume of the reaction
is meson dominated. The mesons, baryons, and antibaryons carry
90\%, 7\%, and 3\% of the total number of particles in the RHIC
cell at $t \geq 10$ fm/$c$ (cf. 85\%, 14.5\%, 0.5\% at SPS and
50\%, 50\%, 0\% at AGS). The microscopic ratios for mesons
(Fig.~\ref{fig5}, right panels) seem to be very close to the SM
ratios. Since the freeze-out occurs at
$t \approx 21$ fm/$c$ in the central cell at RHIC energies, the
matter in the cell is frozen before reaching complete chemical
equilibrium. This circumstance complicates the extraction of the 
chemical and thermal freeze-out parameters by means of the standard 
thermal model fit.

The rapidity distributions of baryon resonances in Au+Au collisions
at $\sqrt{s} = 200$ AGeV are found to be nearly flat in the rapidity 
interval $|y| \leq 3.5$ \cite{resrich}. More than 80\% of the baryon 
non-strange resonances are still $\Delta$'s (1232). The density of 
directly reconstructible baryon resonances, that decay into final 
state hadrons, per unit rapidity at RHIC is quite high, and the
resonance rich matter can be detected experimentally. 
\end{minipage}
\end{figure}

This means that 
not only the temperature, but also the mean mass of a particle is
increased in the cell at RHIC energies, i.e., there should be more
heavy resonances compared with the SPS cell.
Therefore, the ratios of hadronic abundances are studied (see
Fig.~\ref{fig5}). Here the results are presented separately for
non-strange and strange baryons and mesons. In the baryon sector
the resonances dominate over the strange and non-strange baryons
until the end of the simulations. This can be taken as an indication
of the creation of long-lived resonance-rich matter. 
The fraction of baryon resonances is almost 70\% of
all baryons in the cell at RHIC at $5 \leq t \leq 19$ fm/$c$,
while at SPS and AGS the number of baryon resonances decreases from
70\% to 35\%, and from 60\% to 25\%, respectively. The meson
fractions of resonances shrink within the time interval
$5 \leq t \leq 19$ fm/$c$ from 60\% to 30\% (RHIC), 50\% to 20\%
(SPS), and 40\% to 15\% (AGS). But at RHIC energies the 

\vspace*{-0.88cm}
The results of our study can be summarised as follows. 
The formation of long lived resonance-abundant matter is found.
UrQMD predicts that $T = 195 \pm 5$ MeV, $\mu_B = 43 \pm 2$ MeV, 
$150 \leq S/A \leq 170$, and $\langle E \rangle /\langle N_h 
\rangle \approx 1$ GeV at chemical freeze-out in central Au+Au 
collisions at RHIC. The equation of state has a linear dependence 
$P = 0.15\, \varepsilon$.
The UrQMD cell calculations show that strangeness to entropy ratio
monotonically increases with rising $\sqrt{s}$ as
$R_S \cong 0.025$ (AGS), 0.04 (SPS), and 0.05 (RHIC).

\end{document}